\PassOptionsToPackage{table}{xcolor}
\documentclass{optica-article}

\journal{opticajournal} %

\articletype{Research Article}

\usepackage{booktabs}
\usepackage{array}
\usepackage{graphicx}
\usepackage{subcaption}
\usepackage{enumitem}
\usepackage{listings}
\usepackage{hyperref}
\usepackage{float}
\usepackage{siunitx}
\usepackage{tikz,pgfplots,amsmath}
\usetikzlibrary{shapes.arrows, shapes.geometric, arrows.meta, fit, positioning}
\hypersetup{
  colorlinks=true,
  linkcolor=blue!60!black,
  citecolor=blue!60!black,
  urlcolor=blue!60!black,
  pdftitle={A deployed four-city entanglement-based QKD network with real-time key delivery, benchmarked against finite-key security bounds},
  pdfauthor={Gilberto Borges, Saeid Salari, Ekta Panwar, Samgeeth Puliyil, Kartik Kakade, Peter Rapcan, Mario Ziman, Djeylan Aktas}
}

\lstdefinestyle{pythonstyle}{
  language=Python,
  basicstyle=\ttfamily\small,
  keywordstyle=\color{blue!70!black},
  commentstyle=\color{green!40!black},
  stringstyle=\color{red!50!black},
  showstringspaces=false,
  breaklines=true,
  frame=single,
  rulecolor=\color{black!20},
  backgroundcolor=\color{black!2},
  columns=fullflexible
}
\usepackage{arydshln}
\lstdefinestyle{optica}{language=Python,basicstyle=\ttfamily\fontsize{7.5}{9.5}\selectfont,
 keywordstyle=\bfseries\color{blue!50!black}, commentstyle=\color{black!55},
 frame=single,framerule=0.25pt,rulecolor=\color{black!25},
 backgroundcolor=\color{black!3},xleftmargin=4pt,xrightmargin=4pt,
 framesep=4pt,columns=fullflexible,keepspaces=true,showstringspaces=false,
 breaklines=true,aboveskip=5pt,belowskip=6pt,tabsize=4}

\renewcommand{\theequation}{\arabic{equation}}
\renewcommand{\thetable}{\arabic{table}}
\renewcommand{\thefigure}{\arabic{figure}}

\usepackage{cite}
\usepackage{mathtools}

\usepackage{amsthm}

\usepackage{algorithmic}
\usepackage{textcomp}
\usepackage{braket}
\usepackage[normalem]{ulem}

\definecolor{lightgray}{gray}{0.9}
\usepackage{booktabs}
\def\BibTeX{{\rm B\kern-.05em{\sc i\kern-.025em b}\kern-.08em
    T\kern-.1667em\lower.7ex\hbox{E}\kern-.125emX}}

\theoremstyle{definition}
\newtheorem{definition}{Definition}

\begin{document}

\title{A deployed four-city entanglement-based QKD network with real-time key delivery, benchmarked against finite-key security bounds}

\author{Gilberto Borges,\authormark{1,*} Saeid Salari,\authormark{1} Ekta Panwar, \authormark{1} Samgeeth Puliyil, \authormark{1} Kartik Kakade,\authormark{1} Peter Rap\v can, \authormark{1} Mario Ziman, \authormark{1} and Djeylan Aktas\authormark{1,$\dag$}}

\address{\authormark{1}Institute of Physics, Slovak Academy of Sciences, D\'ubravsk\'a cesta 9, 84511, Bratislava, Slovakia}

\email{\authormark{*}gilberto.borges@savba.sk \\ 
\authormark{$\dag$} djeylan.aktas@savba.sk} %

\begin{abstract*} 
Quantum key distribution (QKD) generates shared secret keys whose security rests on the laws of physics rather than on computational assumptions. We report the entanglement-based segment of the Slovak national quantum communication infrastructure, connecting four cities in western Slovakia over deployed telecommunication fiber. Polarization-entangled photons generated at a central node are passively routed by wavelength to form a trusted-node-free, fully meshed sub-network of three cities, with the fourth connected through the central node and all links operated simultaneously. We also introduce a matrix formalism for passive wavelength routing that enables the algorithmic construction of uniquely and fully connected topologies. Unlike most previous demonstrations, which estimate key rates from raw coincidences, all rates reported here are obtained after the complete post-processing chain, with keys delivered in real time via ETSI GS QKD 014. With partial non-local chromatic dispersion compensation, we obtain 13.5~bit/s over 109~km, and 4.3~bit/s over
32.1~dB of accumulated loss between two cities without a direct fiber link. Benchmarking against finite-key security bounds shows that the implemented heuristic privacy amplification can exceed the finite-key bound, and that replacing detect-and-discard with error correction would increase the finite-key rate by a factor of 4.1–6.6, and yield a positive finite-key rate on the 142.9 km link where the implemented scheme yields none.

\end{abstract*}

\section{Introduction}

Quantum networks (QNs) interconnect devices that process, store, and measure quantum information~\cite{kimble2008quantum,wehner2018quantum}. Besides quantum key distribution (QKD)~\cite{scarani2009security,pirandola2020advances}, they enable distributed quantum computing~\cite{main2025distributed} and sensing~\cite{guo2020distributed,malia2022distributed,zhang2021distributed}.

The deployment of QNs has evolved from proof-of-principle experiments at the laboratory or campus scale to field-deployed demonstrations and industry-driven use cases, based on prepare-and-measure, entanglement-based, and continuous-variable QKD protocols, and is boosting the development of commercial QKD solutions. These implementations can be grouped into trusted-node networks, in which key material is relayed hop by hop~\cite{peev2009secoqc,sasaki2011tokyo,stucki2011longterm,wang2014field,dynes2019cambridge,chen2021integrated}, and trusted-node-free networks, in which no intermediate node needs to be trusted~\cite{tang2016mdi,wengerowsky2018entanglement,Joshi_BristolQKD,neumann2022continuous}. Trusted-node architectures allow secure communication over long distances, but the trustworthiness of a node is difficult to guarantee if relay sites are not owned by the network operator. Moreover, this scales poorly with network size, because the number of sites that must be physically secured grows with the topology. In trusted-node-free networks, no intermediate node holds key material, which improves the security of the key at the cost of reduced key rates and a reach limited by photon loss, so that long-range QKD without trusted nodes is not yet possible with current technology~\cite{b3}. This complementarity motivates hybrid architectures, in which trusted-node-free links connect users within a region and trusted nodes bridge the longer distances between regions, thereby limiting the number of nodes that must be trusted.

Among trusted-node-free approaches, entanglement has become a central resource for building multi-user networks. Wengerowsky \emph{et al.}~\cite{wengerowsky2018entanglement} pioneered a fully connected four-user network using wavelength demultiplexing (DEMUX) and multiplexing (MUX) to distribute entanglement, assigning an independent pair of wavelength channels to each pair of users. This approach has since supported QKD in a metropolitan-scale eight-user network~\cite{Joshi_BristolQKD} and the first trans-national entanglement distribution over deployed fiber suitable for QKD~\cite{neumann2022continuous}. Fully connected entanglement-based networks have also been scaled to 40 users by combining wavelength and space-division multiplexing~\cite{liu2022fortyuser}, realized with integrated on-chip photonic sources~\cite{appas2021flexible,wen2022realizing}, and extended to flexible bandwidth allocation~\cite{lingaraju2021adaptive}, time-bin encoding~\cite{fitzke2022scalable}, and energy-time entanglement~\cite{Djeylan150km2016}.

Despite the above-mentioned advances, for QKD to be viable in real-world cryptographic applications, secure key material must be generated at a rate that meets the target application's consumption demands and delivered to the application as it is produced. In many demonstrations, however, the secure key rate is estimated from data processed after the experiment~\cite{neumann2022continuous, Joshi_BristolQKD, wengerowsky2018entanglement}. Real-time key generation and delivery have mostly been reported in trusted-node networks based on prepare-and-measure protocols~\cite{peev2009secoqc,sasaki2011tokyo,dynes2019cambridge,chen2021integrated,ribezzo2023deploying,martin2024madqci}, where standardized key delivery interfaces such as ETSI GS QKD 014~\cite{ETSI_QKD_014} are increasingly adopted~\cite{martin2024madqci}. For entanglement-based QKD, real-time operation has so far been demonstrated mainly over single metropolitan links~\cite{pelet2023operational,selimovic2026deployment,Lecaron_2026}. A recent inter-city entanglement-based network delivered keys in accordance with ETSI GS QKD 014~\cite{dosan2026secure}, with links operated during separate measurement campaigns. To the best of our knowledge, real-time operation of the complete key distillation and delivery chain across a multi-city entanglement-based network, with all links operating simultaneously, has not yet been reported. Moreover, the key produced by such operational pipelines is rarely benchmarked against finite-key security bounds, which is essential to assess whether it can be used in practice.

\begin{figure}[ht]
\centering
\includegraphics[width =\linewidth]{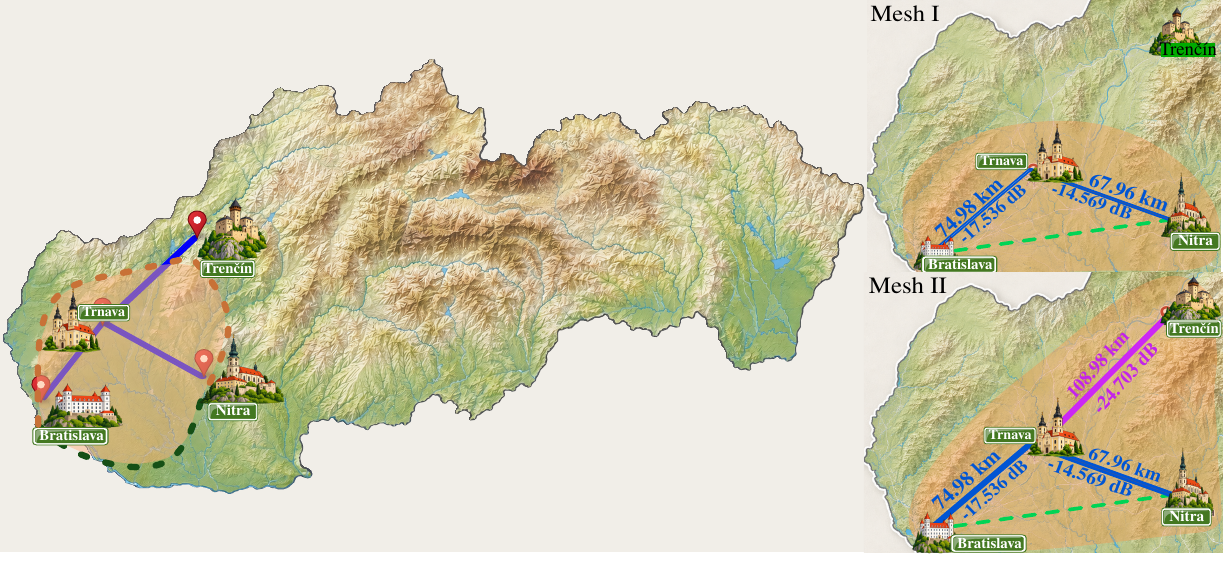}
\caption{\textbf{Topology of the field-deployed entanglement-based quantum network in western Slovakia.} Left: The final network configurations. Solid lines indicate optical fiber links between the central node (Trnava) and the peripheral nodes (Bratislava, Nitra, and Tren\v{c}\'{i}n). Black dashed lines indicate quantum links between nodes without a direct fiber connection, established through our wavelength routing strategy, while orange dashed lines indicate nodes that may be connected through a trusted node. Right: Mesh I, the three-user fully meshed configuration, and Mesh II, the complete deployed network after adding Tren\v{c}\'{i}n, in which all links are operated simultaneously. The color code indicates the wavelength pair of photons that establish the connection, while the distances were measured using an OTDR.}
\label{fig:network_slovak_map}
\end{figure}

Here, we demonstrate the feasibility of a four-node inter-city entanglement-based quantum network that continuously distills cryptographic keys for consumption via the ETSI GS QKD 014 interface. The network is part of the Slovak Quantum Communication Infrastructure (SkQCI), which belongs to the European Quantum Communication Infrastructure (EuroQCI) launched in 2019~\cite{euroqci}, and connects Bratislava, Trnava, Nitra, and Tren\v{c}\'{i}n in western Slovakia (Figure~\ref{fig:network_slovak_map}). A single broadband entangled-photon-pair source at the central node in Trnava feeds a quantum network provider (QNP), which routes conjugate-wavelength channels to the nodes. In Mesh I, Trnava, Bratislava, and Nitra form a fully meshed, trust-node-free sub-network. In Mesh II, Tren\v{c}\'{i}n, the farthest node from Trnava, is additionally connected to the central node, and all links are operated simultaneously. Notably, all links demonstrated here are trusted-node-free. In principle, the network could be further extended to full four-node connectivity by employing Trnava as a trusted node for key relay, connecting Bratislava and Nitra to Tren\v{c}ín, resulting in a hybrid architecture.

This implementation contributes to the technological maturity of QKD by revealing constraints that are less relevant to the asymptotic estimation of the key rate. Real-time key distillation requires shorter integration times, so each link must sustain a key rate high enough for real-world use. Given the distances involved, wavelength channels become a precious resource, since assigning too many channels to a single node increases noise, while splitting the channels too much degrades the performance of long links. Existing DEMUX/MUX architectures either require a number of wavelength channels that grows quadratically with the number of users~\cite{wengerowsky2018entanglement}, or split all channels with beam splitters (BSs) whose loss grows with the number of users and which create \textit{premium links}, in which some pairs of users share more than one channel~\cite{Joshi_BristolQKD}. Premium links waste entanglement resources that could serve other pairs and hinder the automatic identification of channel time delays, since two channels between the same pair produce two coincidence peaks. 

To address these limitations, we describe the passive wavelength assignment of the QNP via a matrix that directly encodes user connectivity. This allows us to verify and enumerate uniquely connected fully meshed topologies, in which each pair of users shares exactly one entangled wavelength pair. In this way, four users are fully connected via three wavelength pairs, half of those required by dedicated channels ~\cite{wengerowsky2018entanglement}, without premium links and using only 1×2 splitters on the signal channels, as discussed in Sec.~\ref{subsec:connectivity_matrix}. This formalism underpins the exact resource laws for passive wavelength routing derived in our companion work ~\cite{panwar2026exact}. We validate the architecture on a four-node laboratory testbed and on the deployed network.

In the field, we optimize the coincidence time window, analyze the effect of partial non-local chromatic dispersion compensation, and quantify the noise introduced when a new user is added to an already running connection. 

\section{Methods} \label{sec:Methods}
The experimental setup of the networks implemented in this work are illustrated in Figure~\ref{fig:experiment_schematic}, and it is composed of three main subsystems: (i) a polarization-entangled photon-pair source, (ii) a quantum network provider (QNP) equipped with a dynamic polarization controller (PolCo) module to pre-compensate for fiber propagation drift, and (iii) polarization analysis modules (PAM) representing the network nodes.
\subsection{Entanglement generation and chromatic dispersion influence} \label{subsec:Entanglement_generation}
Entanglement distribution within this architecture relies on spectral multiplexing. This approach exploits the frequency correlations of photon pairs generated through spontaneous parametric down-conversion (SPDC)~\cite{mandel_wolf_1995, etb_review}. 

\begin{figure}[h]
\centerline{\includegraphics[width=\textwidth]{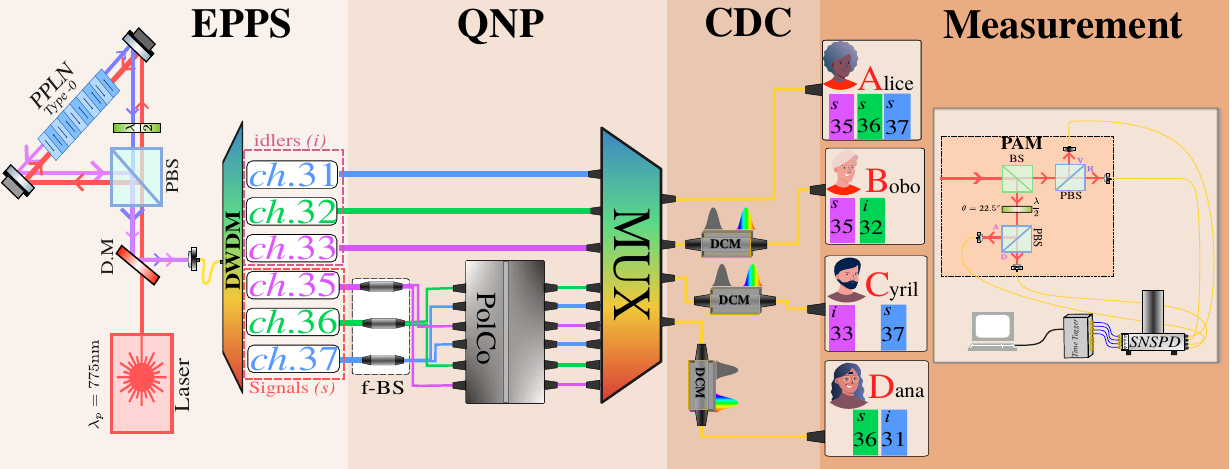}}
\caption{\textbf{Experimental schematic for the entanglement generation, distribution, and measurement} The EPPS denotes an entangled photon pair source. The generated spectrum is then distributed by the Quantum Network Provider (QNP), which features a polarization controller (PolCo). In the QNP output, chromatic dispersion compensation (CDC) is performed. The Polarization Analyses Module (PAM), passively performs the measurements necessary to implement the BBM92 protocol.}
\label{fig:experiment_schematic}
\end{figure}

As illustrated in Fig.~\ref{fig:experiment_schematic}(b), our specific entangled photon-pair source (EPPS) utilizes a type-0 SPDC QPM regime, where the polarizations of the pump beam and the down-converted photons are identical. The quasi-phase-matched PPLN crystal is embedded within a Sagnac interferometer and pumped bidirectionally by a continuous-wave laser at $\lambda_{p} \approx 775 \: \text{nm}$~\cite{b2,li2015}. This configuration yields the polarization-entangled state:

\begin{equation} \label{eq:polarization_entangled_state}
    \ket{\Psi} = \frac{1}{\sqrt{2}} \left( \ket{H_{\omega_{i}}, H_{\omega_{s}}} + e^{i\phi} \ket{V_{\omega_{i}}, V_{\omega_{s}}}  \right) \:,
\end{equation}
where $H$ ($V$) denotes the horizontal (vertical) polarization state, and $\phi$ is the phase corresponding to the coherent superposition of the two bidirectional generation processes. While $\omega_{s}$ and $\omega_{i}$ are the frequencies of the two generated photons, which are related to the pump laser frequency by $\omega_{s} + \omega_{i} = \omega_{p}$, due to energy conservation in the SPDC process.

To distribute the entangled photon pairs among the network nodes, the generated spectrum is demultiplexed using a $100~\si{\giga\hertz}$ ITU-grid DWDM. After discarding the channel whose central wavelength corresponds to spectrally degenerate photons, we select $M/2$ consecutive ITU channels on each side of it, retaining both members of each correlated pair. The selected channels, indexed in spectral order and represented by their central wavelengths $(\lambda_1,\dots,\lambda_M)$, are distributed among $N$ users $(u_1,\dots,u_N)$. For a given idler wavelength index $i \leq M/2$, the corresponding signal index $s$ of the entangled pair $(\lambda_i,\lambda_s)$ is defined as:

\begin{equation} \label{eq:simetric_wavelengths}
    s = M - i + 1 \:.
\end{equation}

Chromatic dispersion (CD) ~\cite{saleh2019fundamentals}, typically $17~\si{ps/(nm\cdot km)}$ in standard fiber at $1550~\si{nm}$~\cite{itu-t-g652-2024}, causes the broadening of the coincidence peaks. If one of the photons, namely the idler, is sent through a normal dispersive medium such that its accumulated dispersion satisfies $\beta_{i} = - \beta_{s}$, the total dispersion of the biphoton wavepacket becomes zero. This effect is known as non-local chromatic dispersion cancellation~\cite{Franson1992,neumann2021nonlocal}. Here we perform partial dispersion compensation by using fixed dispersion compensation modules (DCMs) with a dispersion of $-1260~\si{ps/nm}$, equivalent to $74.1~\si{km}$ of standard fiber, placed at the central node, as shown in Fig.~\ref{fig:experiment_schematic}.

\subsection{Quantum Network provider} \label{QNP_description}
Following the DEMUX stage at the EPPS output, the generated photon pairs are routed through the QNP. In our implemented architecture, the signal photons corresponding to ITU channels 35, 36, and 37 are individually split using standard 1×2 beam splitters and then routed to the polarization control (PolCo) module.

The PolCo module, manufactured by Quantum Optics Jena GmbH, is a 6-channel device that utilizes electronically driven liquid crystal cells to independently transform the polarization state of each port. This module dynamically pre-compensates for polarization drifts induced by fiber propagation. This is achieved via a continuous feedback loop that monitors the QBER and supplies real-time electrical correction signals to the PolCo to minimize it.

In contrast to the split signal photons, the idler photons (ITU channels 31, 32, and 33) are routed directly from the source to the multiplexing (MUX) module. Here, they are recombined with the corresponding split signal photons using a cascade of single-channel DWDM add-drop filters, as schematically illustrated in Fig.~\ref{fig:experiment_schematic}(a). The resulting four output fibers from the MUX module constitute the final QNP outputs, which are then distributed to the respective network nodes. 

\subsection{Polarization analyses modules} \label{PAM_nodes_description}

The four output fibers of the QNP serve as inputs for the user measurement modules, which represent the distributed network nodes. As depicted in Figure~\ref{fig:schematic_post_processing_qkd}, each node comprises a Polarization Analysis Module (PAM), a cryostat housing four Superconducting Nanowire Single-Photon Detectors (SNSPDs), and a Swabian Instruments Time Tagger synchronized to an external $100 \: \text{MHz}$ clock. The Time Tagger assigns a precise timestamp to every detection event registered by the SNSPDs.

The PAMs, manufactured by Quantum Optics Jena GmbH, are passive, fiber-based optical devices, and their schematic representation of their free-space equivalent is provided in Figure~\ref{fig:experiment_schematic} (c).  By analyzing the Time Tagger's output buffer, it is possible to seamlessly infer both the arrival time and the projected polarization state of each detected photon

Each node is equipped with an internal server that hosts the control software for key management and the PolCo feedback loop. One of the PAM clocks is designated as the reference clock, ensuring that all Time Taggers across the network are strictly synchronized to a single, unified global time standard.

\subsection{QKD post-processing pipeline and finite-key effects} \label{subsec:post-processing_software}

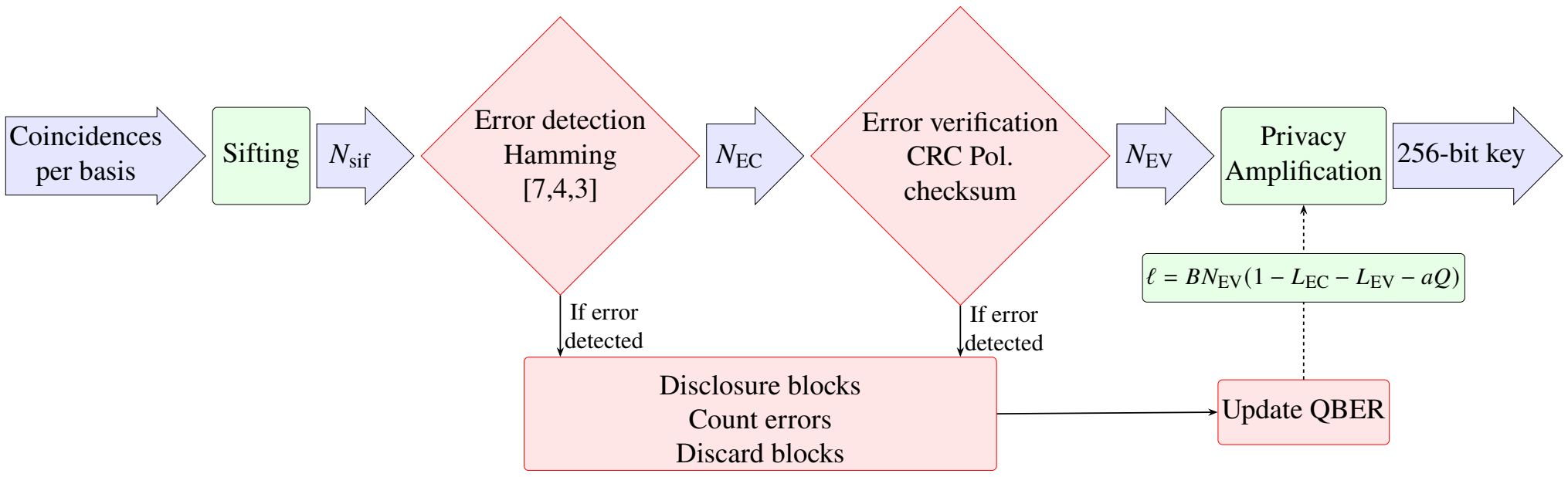
\begin{figure}[h]
\resizebox{\textwidth}{!}{%
\begin{tikzpicture}[node distance=1.5cm and 0.2cm]
		\node[single arrow, draw, fill=blue!10, shape border rotate=0, minimum height=1cm, minimum width=3cm, align=center,font=\Huge] (arrow1) {Coincidences \\ per basis};
		
		\node[draw, rounded corners=4pt, fill=green!10, right=of arrow1, minimum height=3cm, minimum width=3cm,font=\Huge] (sifting) {Sifting};
		
		\node[single arrow, draw, fill=blue!10, shape border rotate=0, minimum height=3cm, minimum width=3cm, text centered,right=of sifting] (arrow2) {\Huge $N_\text{sif}$};
		
		\node[diamond, draw=red, fill=red!10, aspect=1, minimum width=3cm, align=center,right=of arrow2,font=\Huge] (error1) {Error detection \\ Hamming\\ {[7,4,3]}};
		
		\node[single arrow, draw, fill=blue!10, shape border rotate=0, minimum height=3cm, minimum width=3cm, text centered,right=of error1] (arrow3) {\Huge $N_\text{EC}$};
		
		\node[diamond, draw=red, fill=red!10, aspect=1, minimum width=3cm, align=center,right=of arrow3,font=\Huge] (error2) {Error verification \\ CRC Pol.\\ checksum};
		
		\node[single arrow, draw, fill=blue!10, shape border rotate=0, minimum height=3cm, minimum width=3cm, text centered,right=of error2,font=\Huge] (arrow4) {$N_\text{EV}$};
		
		\node[draw, rounded corners=4pt, fill=green!10, right=of arrow4, minimum height=3cm, minimum width=3cm,align=center,font=\Huge] (privacy) {Privacy \\ Amplification};
		
		\node[single arrow, draw, fill=blue!10, shape border rotate=0, minimum height=3cm, minimum width=3cm, text centered,right=of privacy] (arrow5) {\Huge 256-bit key};
		
		\coordinate [below=1.5cm of error1.south,xshift=-1cm] (left-anchor);
		\coordinate [below=1.5cm of error2.south,xshift=1cm] (right-anchor);
		
		\node[rounded corners=4pt, draw=red, fill=red!10, below=2cm of error1, minimum height=3.5cm, minimum width=5cm,align=center,fit=(left-anchor) (right-anchor),font=\Huge] (blocks) {Disclosure blocks \\ Count errors\\ Discard blocks};
		
		\draw [-Stealth,align=center,very thick] (error1.south) -- node[right,font=\huge] {If error\\ detected} (error1.south |- blocks.north);
		\draw [-Stealth,align=center,very thick] (error2.south) -- node[right,font=\huge,yshift=0.1cm] {If error\\ detected} (error2.south |- blocks.north);
		
		\node[draw=red, fill=red!10, rounded corners=4pt, below=5.4cm of privacy, minimum height=2cm, minimum width=3cm,align=center,font=\Huge] (qber) {Update QBER};
		
		\draw [-Stealth,align=center,very thick] (blocks.east) -- (qber.west);
		
		\draw [-Stealth,align=center,dashed,very thick] (qber.north) -- (privacy.south);
		
		\node[draw, rounded corners=4pt, fill=green!10, below=1.5cm of privacy, minimum height=1.5cm, minimum width=3cm,align=center] (eqn) {\huge$\ell = B N_\text{EV}(1-L_\text{EC}-L_\text{EV}-a Q)$};
		
	\end{tikzpicture}}
\caption{Schematic diagram for the QKD post-processing software for the evaluation system used in this work. For more details please refer to the main text.}
\label{fig:schematic_post_processing_qkd}
\end{figure}

The distillation of the final 256-bit symmetric keys from the coincidence counts is handled by a custom software suite developed by Quantum Optics Jena GmbH ~\cite{QOJ_error_det_pat}. This post-processing pipeline avoids active bit-flipping during key reconciliation to reduce required computational resources and maximize processing speed. The distilled keys are stored in pools with a capacity of 30,000 keys per link. These keys are available for consumption through the ETSI GS QKD 014 API standard~\cite{ETSI_QKD_014}. The operational pipeline of this routine is detailed below, and a summarizing diagram is provided in Fig.~\ref{fig:schematic_post_processing_qkd}.

After compensating for the time delay in the synchronization stage, raw timestamps are filtered to identify simultaneous detection events that fall within a predefined coincidence time window, which is then sifted, giving the number of bits measured in the same basis ($N_{\mathrm{Sif}}$). Next, the sifted bitstream is divided into consecutive 7-bit blocks, and instead of employing error-correction protocols such as Cascade~\cite{Brassard1994} or low-density parity-check (LDPC) codes~\cite{Elkouss2009}, the system uses a block-based error-detection strategy, an error-cleaning (EC) step based on the binary Hamming $[7,4,3]$ code~\cite{hamming1950,macwilliams1977, QOJ_error_det_pat}. This way, both parties compute and exchange the 3-bit syndrome of each block. Since the code has minimum distance 3, all 1- and 2-bit errors in a block are detected. Blocks whose syndromes differ are disclosed, so that their errors can be counted to estimate the quantum bit error rate (QBER), and then discarded. The syndromes of the kept blocks disclose $L_{EC} = 3/7$ bits per kept bit.

To verify the integrity of the remaining string, an error verification (EV) step groups the surviving bits into 256-bit frames. A 16-bit cyclic redundancy check (CRC-16)~\cite{Peterson1961} is computed for each frame and the tags are exchanged. Frames whose tags differ are disclosed to count their errors and update the QBER, and then discarded. Since each kept frame discloses its 16-bit tag, the EV leakage is $L_{EV} = 16/256$ bits per kept bit.

Privacy amplification (PA) compresses the verified key with a cryptographic hash function of the SHA family~\cite{carter1979}. The output length is set from the QBER, the information disclosed in the EC and EV stages, and a balance factor $B$ that penalizes unequal coincidence counts. With $c_i$ the coincidence counts of the correlated detector pairs, $i \in \{HH, VV, DD, AA\}$, $B$ is given by:
\begin{equation} \label{eq:unbalancing_factor_PA}
    B = 4 \frac{ \min_{i} c_{i}}{\sum_{i} c_{i}}\,,
\end{equation}
so that $B = 1$ for perfectly balanced counts. The final key length is empirically estimated as:

\begin{equation} \label{eq:QOJ_key_bound_model}
    \ell = B\, N_{EV} \left(1 - L_{EC} - L_{EV} - aQ\right),
\end{equation}
where $N_{EV}$ is the number of bits surviving EV (all bits of the kept blocks), $Q$ is the QBER combined over both bases, and $aQ$ is a linear estimate of the eavesdropper's information per key bit, with the slope $a$ being a design choice of the implementation, whose value we inferred from the experimental data. Since the linear-decreasing model used to estimate the eavesdropper information does not follow from a security proof, it may fail to capture the potential information leakage arising from a biased source or from detector imbalance, and the balanced PA can be viewed as a heuristic correction term in the SKR.

The standard entropic key-rate considering the finite key size is given by:

\begin{equation} \label{eq:finite_key_H(e)}
    \ell_{f} \le N_{\mathrm{sif}}\left[1 - h(Q + \mu)\right] - I_\mathrm{leak}
         - 2\log_{2}\frac{1}{2\varepsilon_{\mathrm{PA}}}
         - \log_{2}\frac{2}{\varepsilon_{\mathrm{cor}}},
\end{equation}
where $N_{\mathrm{sif}}$ is the number of sifted bits from both bases in a privacy-amplification block, $h(x) = -x\log_2 x - (1-x)\log_2(1-x)$ is the binary entropy, and $Q$ is the QBER combined over both bases, which bounds the phase error rate for a balanced basis choice. The term $\mu$ accounts for the statistical fluctuation of the phase-error estimate~\cite{tomamichel2012tight}. The information disclosed during reconciliation is $I_\mathrm{leak} = f N_{\mathrm{sif}}\, h(Q)$, where $f \geq 1$ is the reconciliation inefficiency, which for optimized LDPC and Cascade implementations reaches $f \approx 1.05$--$1.2$, while short blocks or simpler protocols such as Winnow give values up to about 1.5~\cite{Brassard1994,buttler2003winnow,Elkouss2009,martinezmateo2015cascade}. For the implemented EC and EV, $N_{\mathrm{sif}}$ is replaced by $N_{EV}$ and $I_\mathrm{leak} = N_{EV}(L_{EC} + L_{EV})$, and the last term is dropped, since the CRC tags are already included in $L_{EV}$. The last two terms are the costs of privacy amplification and error verification, where $\varepsilon_{\mathrm{sec}} = 2\varepsilon_{\mathrm{PE}} + \varepsilon_{\mathrm{PA}}$ and $\varepsilon_{\mathrm{cor}}$ are the secrecy and correctness parameters, set here to the commonly adopted values $\varepsilon_{\mathrm{sec}} = 10^{-10}$ and $\varepsilon_{\mathrm{cor}} = 10^{-15}$~\cite{lim2014concise}. In the asymptotic case, the key rate is given by:
\begin{equation} \label{eq:asymptotic_key}
    R_{\infty} = n_{\mathrm{sif}}\left[1 - h(Q) - f\,h(Q)\right],
\end{equation}
where $n_{\mathrm{sif}}$ is the sifted bit rate, and for the implemented EC and EV $n_{\mathrm{sif}}$ is replaced by $n_{EV}$ and $f\,h(Q)$ by $L_{EC} + L_{EV}$.

The comparison of the experimental data obtained in this work with the key rate computed by the implemented solution (Eq.~\ref{eq:QOJ_key_bound_model}) and with the secure key bounds predicted by Eqs.~\ref{eq:finite_key_H(e)} and~\ref{eq:asymptotic_key} gives interesting insights about the optimization of practical QKD networks, as will be discussed in Section~\ref{sec:Results}.

\subsection{Matrix formalism for passive wavelength routing and uniquely connected topologies} \label{subsec:connectivity_matrix}
Unlike treatments of entanglement distribution over
random network topologies~\cite{romancino2026percolation} or spectrum allocation in flexible-grid networks~\cite{alnas2022optimal}, we address the deterministic design of a given passive QNP. 

We describe the wavelength assignment of a network with $M$ wavelength channels and $N$ users by the $M \times N$ QNP matrix $\mathbf{Q} = [Q_{m,n}]$, where $Q_{m,n} = 1$ if user $u_n$ is assigned wavelength $\lambda_m$, and $Q_{m,n} = 0$ otherwise. Each row of $\mathbf{Q}$ corresponds to a wavelength and each column to a user, as in the examples of Tables~S1 and~S2. According to Eq.~\eqref{eq:simetric_wavelengths}, rows $m$ and $M+1-m$ correspond to the two photons of an entangled pair, so two users share a connection whenever their columns contain ones at such symmetric positions. The number of connections between users $u_n$ and $u_{n'}$ is given by the elements of the $N \times N$ connectivity matrix
\begin{equation} \label{eq:connectivity_matrix}
    \mathbf{C} = \mathbf{Q}^{\mathrm{T}} \mathbf{R}_{M} \mathbf{Q}\:, \qquad C_{n,n'} = \sum_{m=1}^{M} Q_{m,n}\, Q_{M+1-m,n'}\:,
\end{equation}
where $\mathbf{R}_M$ is the $M \times M$ reversal (exchange) matrix, with ones on the anti-diagonal and zeros elsewhere.  Sandwiching $\mathbf{R}_M$ by $\mathbf{Q}$ reverses its rows, aligning each wavelength with its entangled partner. Each product $Q_{m,n} Q_{M+1-m,n'}$ is therefore equal to 1 only if $u_n$ receives $\lambda_m$ and $u_{n'}$ receives its partner $\lambda_{M+1-m}$, and the sum counts the entangled pairs shared by the two users. Since $\mathbf{R}_M$ is symmetric, $\mathbf{C} = \mathbf{C}^{\mathrm{T}}$, which reflects the bidirectional nature of the entangled links.

A user that receives both photons from a considered pair cannot use it to establish a link, so the pair is wasted. Each such pair contributes to the diagonal of $\mathbf{C}$, in particular the corresponding element reads $C_{n,n}=2$. Since all elements of $\mathbf{C}$ are non-negative, a network free of self-connections must satisfy
\begin{equation} \label{eq:trace_constraint}
    \mathrm{Tr}(\mathbf{C}) = \sum_{n=1}^{N} C_{n,n} = 0 \:.
\end{equation}

\subsubsection{Uniquely connected users}

\begin{definition}[Uniquely connected users]
    Two users $u_{n}$ and $u_{n'}$ are uniquely connected (UC) if they share exactly one entangled wavelength pair, \textit{i.e.}, if $C_{n,n'} = 1$.
\end{definition}

A fully meshed UC network free of self-connections is therefore characterized by
\begin{equation} \label{eq:UC_condition}
    \mathbf{C} = \mathbf{J}_N - \mathbf{I}_N \:,
\end{equation}
where $\mathbf{J}_N$ is the $N \times N$ all-ones matrix and $\mathbf{I}_N$ is the identity matrix. This condition can be verified directly for any QNP matrix. Based on it, we developed an algorithm that, given the total number of wavelengths, the subset of channels to split, and the number of users, returns all valid QNP matrices that satisfy the constraints of this work. Further details on the code are provided in Supplement 1, along with examples of how to apply this formalism to describe a given network topology.

All configurations implemented in this work were designed and verified with
this framework to be uniquely connected, as shown in Sec.~\ref{subsec:network_architectures}.
\subsection{Implemented network architectures}\label{subsec:network_architectures}
\begin{figure}[h]
    \centering
    \includegraphics[width=\linewidth]{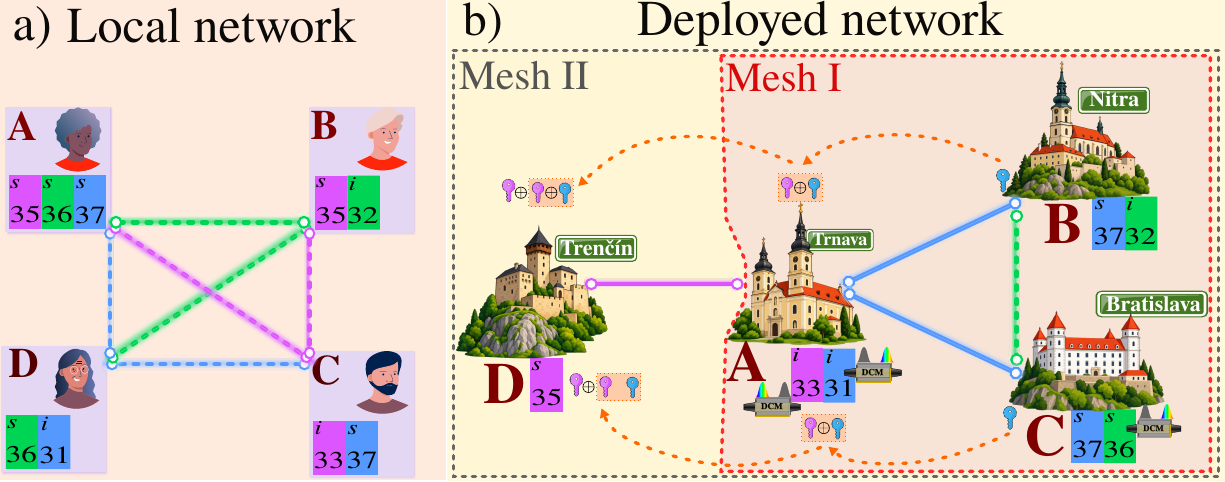}
    \caption{\textbf{Topological diagrams.} (a) Local four-node testbed implementing a four users fully meshed network. (b) Mesh~I, three-user fully meshed configuration (Trnava, Bratislava, Nitra), and Mesh~II, after adding Tren\v{c}\'in and the dispersion compensation modules (DCM).  Orange dashed arrows indicate the possibility of having secure communication with A being a trusted node. Boxed numbers give the ITU channel delivered to each node.}
    \label{fig:topo_ABC}
\end{figure}

To investigate various aspects of the proposed network architecture, we conducted measurements on both a local test network and a field-deployed network. Here, we report results from different topologies to explicitly determine the impact of parameters such as entanglement distribution strategy, chromatic dispersion compensation, and the QKD post-processing software pipeline on the performance of continuously running QKD. The main architectures we have employed here are summarized in Fig.~\ref{fig:topo_ABC}, where the ITU channels routed to each user are represented by the numbers in the grey squares, and the correlated spectral pairs are highlighted by the color code. The lines indicate the links between nodes, with the color indicating which wavelength-channel pair establishes the link. We also distinguish between two cases for the links between nodes: a solid-line arrow indicates a direct optical fiber connection carrying photons between the nodes, while a dashed-line arrow indicates the absence of such a direct connection. The dashed orange line indicates that a key forward strategy, with A being a trusted node, could establish secure communication for C--D and B--D.

The network configurations corresponding to Mesh I and II, Figure~\ref{fig:topo_ABC}b, represent the intercity-deployed network in a star topology, with the source and the QNP located at the central node in Trnava. In the following section, we discuss in more detail how these configurations were useful for providing insights into the performance optimization of the field-deployed QKD system.

\section{Results} \label{sec:Results}

\subsection{Pre-deployment evaluation of the entanglement-based network} \label{subsec:local_entangl_test}
\begin{figure}[t]
    \centering

    \begin{subfigure}[b]{0.48\textwidth}
        \centering
        \includegraphics[width=\linewidth]{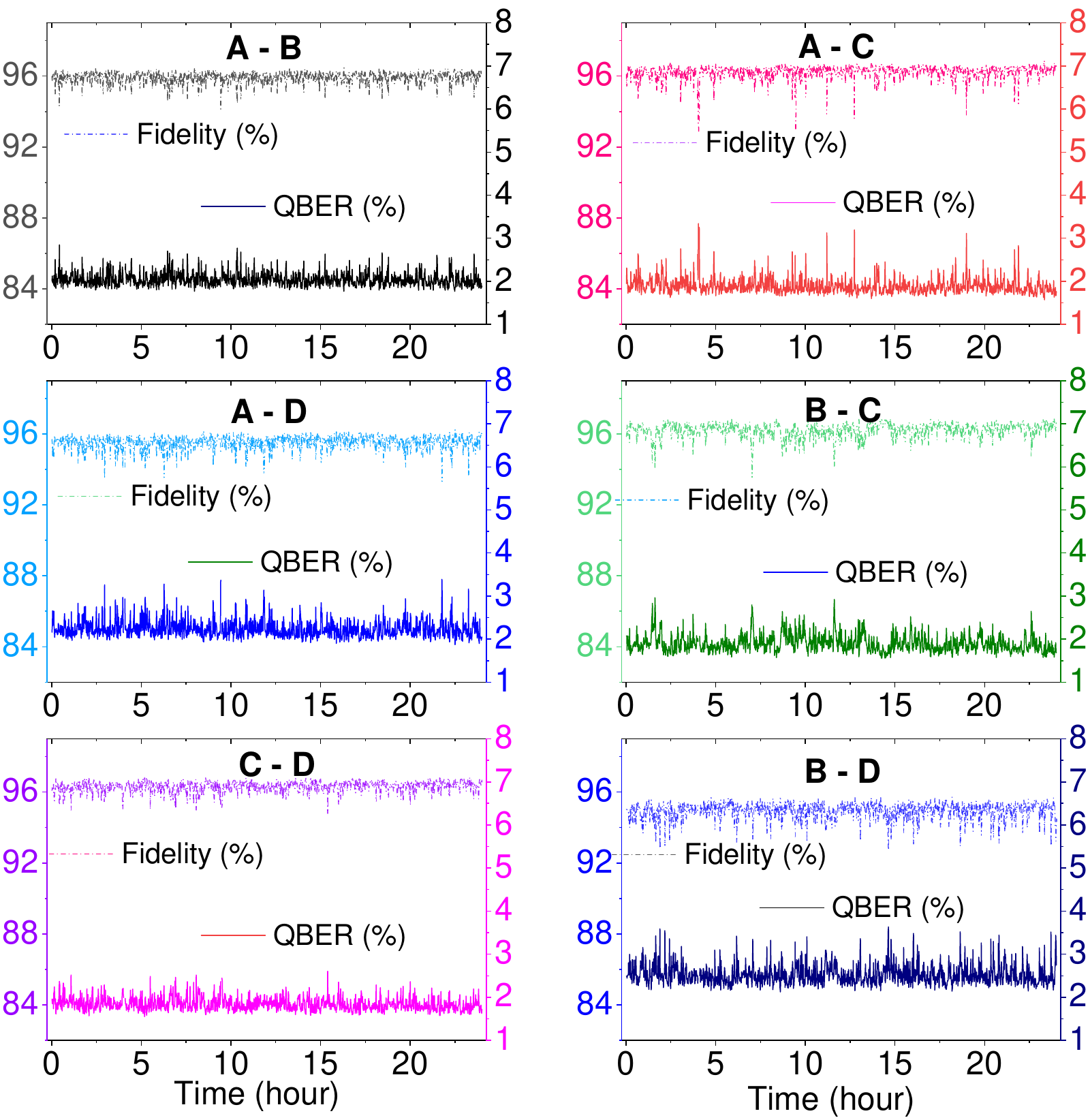}
        \caption{Fidelity and QBER}
        \label{fig:qber_fidelity_local_network}
    \end{subfigure}
    \hfill
    \begin{subfigure}[b]{0.48\textwidth}
        \centering
        \includegraphics[width=\linewidth]{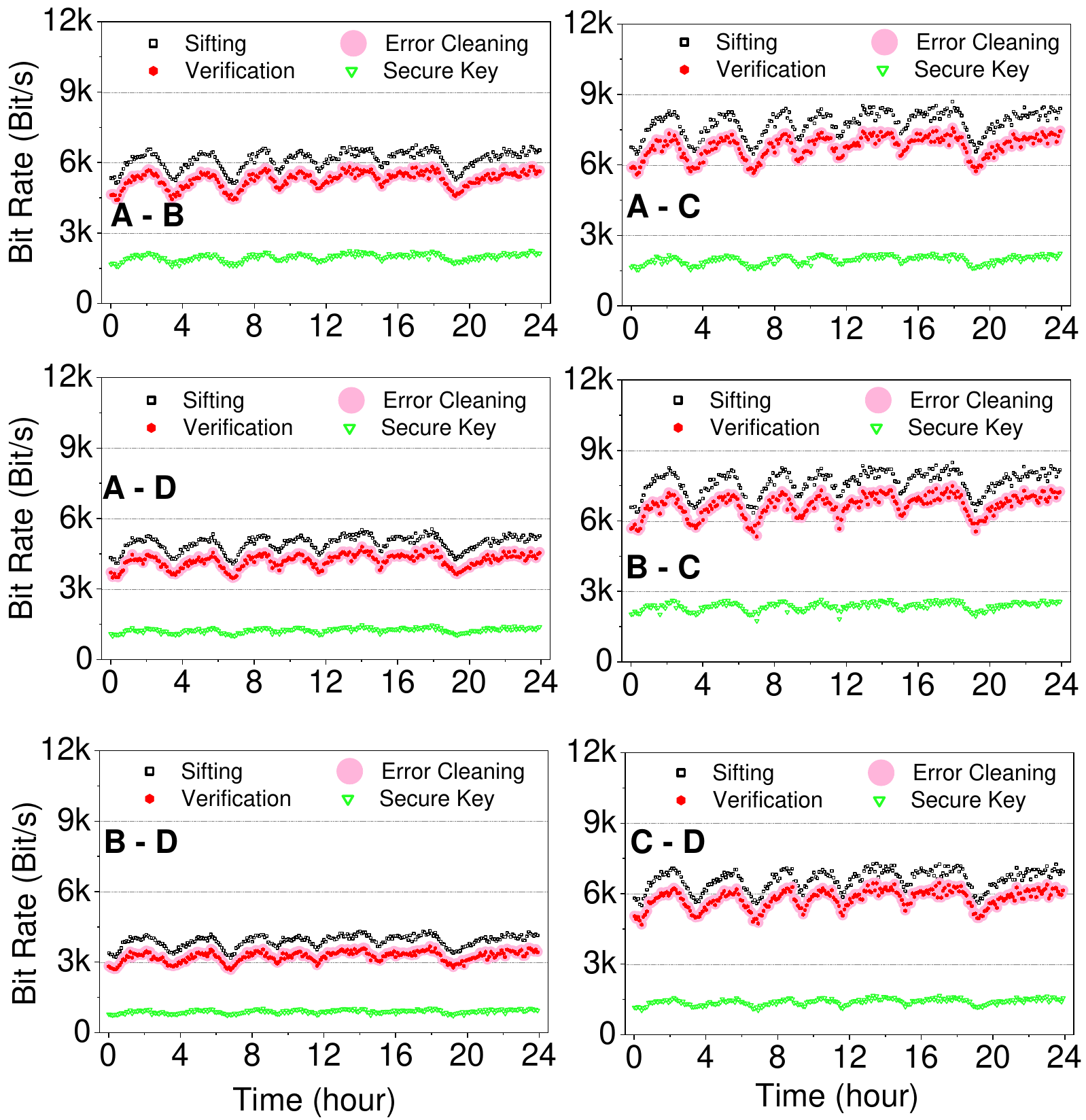}
        \caption{Secure key rates}
        \label{fig:skr_local_network}
    \end{subfigure}

    \caption{Results obtained over a 24-hour integration period in the pre-deployment stage for a four-node fully connected network. (a) Fidelity and QBER. (b) Sifted, error-verified, and final key rates.}
    \label{fig:local_test_results}
\end{figure}

During the pre-deployment stage, we implemented a four-node fully connected network and tested the resilience of the network hardware under unstable conditions, with a total detected coincidence rate of $\approx \qty{7.4e4}{\per\second}$. Here, we employed a custom-built entangled photon-pair source (EPPS) in the same experimental schematic shown in Figure~\ref{fig:experiment_schematic}, and reproduced the network topology of Figure~\ref{fig:topo_ABC}a. The reported performance metrics here are shown for a representative 24-hour period of continuous operation without manual laser drift compensation. Following the temporal synchronization and coincidence filtering (detailed in section~\ref{sec:Methods}), we extracted the experimental figures of merit for our implemented strategy: the fidelity of the prepared state with respect to the maximally entangled target state $\ket{\Phi^{+}}=\frac{1}{\sqrt{2}}(\ket{++} + \ket{--})$, the quantum bit error rate (QBER) and the key rate (KR). These results are summarized in Fig.~\ref{fig:local_test_results}.

Despite the intensity instability of the pump laser, the fidelity ($f$) of the quantum state remains constant over time, as shown in Figure~\ref{fig:qber_fidelity_local_network}. Because the measured QBER closely tracks the state fidelity for the source used here~\cite{Gisin2002}, we correspondingly observe a highly stable QBER. This indicates that the physical hardware layer does not significantly degrade the quality of the generated, propagated, and measured quantum states. As a consequence of this robustness against source instability, we observe a significant volume of secure bits being distilled per second, as shown in Figure~\ref{fig:skr_local_network}.

In addition to hardware performance, we could benchmark the post-processing strategy (as detailed in Section~\ref{subsec:post-processing_software}) in the low-QBER, high-coincidence-count regime.

The slope $a$ was determined by fitting $\ell / (B N_{EV}) = 1 - L_{EC} - L_{EV} - aQ$ to the reported key rates of each one-minute sample. This gives $a = 4.99$, implying that according to Equation~\ref{eq:QOJ_key_bound_model} the key rate goes to zero for a QBER value of $\approx 10.2\%$.

\begin{figure} [h]
    \centering
    \includegraphics[width=\linewidth]{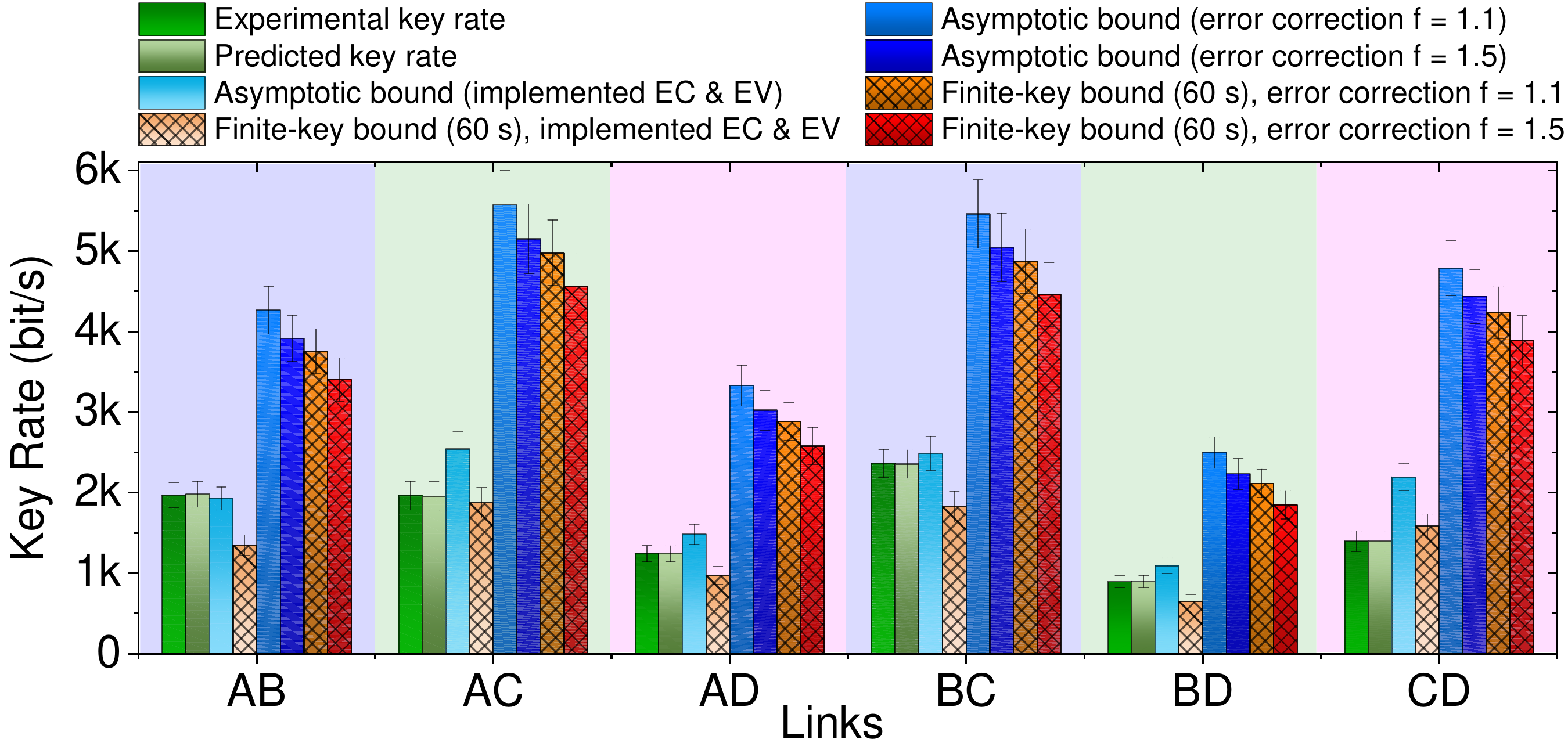}
    \caption{Secret key rate (SKR) of the six links of the four-node network, averaged over 24~h of 1-minute samples. Error bars show the standard deviation between samples. For each link, the experimental SKR is shown together with the predicted key rate from Eq.~\eqref{eq:QOJ_key_bound_model}. These are compared with security bounds computed from the per-basis coincidence data, for the implemented error cleaning and verification (EC \& EV, green) and for conventional error correction with inefficiency $f = 1.1$ and $1.5$ (blue). Solid bars are asymptotic bounds; hatched bars are finite-key bounds for 60~s privacy-amplification blocks ($\varepsilon_{\mathrm{sec}} = 10^{-10}$, plus $\varepsilon_{\mathrm{cor}} = 10^{-15}$ for error correction).}
\label{fig:keyrate_comparsions}
\end{figure}

We then compare the experimental key rate obtained with this evaluation system with the predictions of Eqs.~\ref{eq:finite_key_H(e)} and~\ref{eq:asymptotic_key}. These results are summarized in Figure~\ref{fig:keyrate_comparsions}, in which the asymptotic and finite-key bounds are obtained by using the experimental values in Eqs.~\ref{eq:asymptotic_key} and~\ref{eq:finite_key_H(e)}, respectively, with the finite-key bounds calculated for PA blocks of $60~\si{\second}$. The distilled key rate exceeds the finite-key bound of the implemented EC and EV on five of the six links, by a factor of up to 1.46, although on A--C the excess lies within one standard deviation. The exception is C--D, where the balance factor $B$ strongly reduces the key. This excess is mostly due to the absence of finite-size corrections in Eq.~\ref{eq:QOJ_key_bound_model}, since for $60 \si{\second}$ blocks the statistical correction $\mu$ is comparable to the QBER itself, while the use of a linear function instead of the binary entropy increases the key rate by about $10\%$. Replacing the detect-and-discard strategy with error correction would increase the finite-key bound by a factor of 2.4--3.3 (for $f = 1.5$--$1.1$), since error correction discloses less
information ($f\,h(Q) \approx 0.16$--$0.21$ bits per sifted bit, against $L_{EC} + L_{EV} \approx 0.49$) and does not discard the correct bits of the blocks containing errors.

\subsection{Field-deployed entanglement-based QKD network} \label{subsec:field_deployed_network}
While pre-deployment tests provided essential operational baselines, field deployment introduces practical challenges. Most performance limitations in this regime are governed by the signal-to-noise ratio (SNR), which, for entanglement-based photonic implementations, is strictly tied to the coincidence-to-accidental ratio (CAR)~\cite{Neumann2021Model}.

In our deployed architecture, a stable mono-block EPPS developed by Zero-Third GmbH and the QNP are located at the central node A in Trnava. We characterized the optical fiber links using an optical time-domain reflectometer (OTDR, Yokogawa AQ7280). In Figure~\ref{fig:network_slovak_map}, we show the total losses and lengths measured at $1550~\si{\nano\meter}$. All links exhibit average attenuation rates below the maximum of $0.30~\si{\decibel\per\kilo\meter}$ specified by ITU-T G.652.D for the 1530–1565 nm range~\cite{itu-t-g652-2024}.

Initially, we operated without chromatic dispersion compensation (CDC), targeting a fully meshed three-user network comprising nodes A, B, and C (Trnava, Nitra, and Bratislava, respectively), as shown in Fig.~\ref{fig:topo_ABC}b. To quantify the impact of adding a user to an established two-user link, we first assigned the correlated ITU channel pair $(32,~36)$ to B and C. We then passively split ITU channel 37 equally between B and C and assigned its correlated channel, ITU 31, to A, completing the Mesh I topology in Fig.~\ref{fig:topo_ABC}b. This way, the photons from channel 37 at B and C act as noise on the existing link between B and C.

\begin{figure}[htpb]
    \centering

    \begin{subfigure}[t]{0.48\textwidth}
        \centering
        \includegraphics[width=\linewidth]{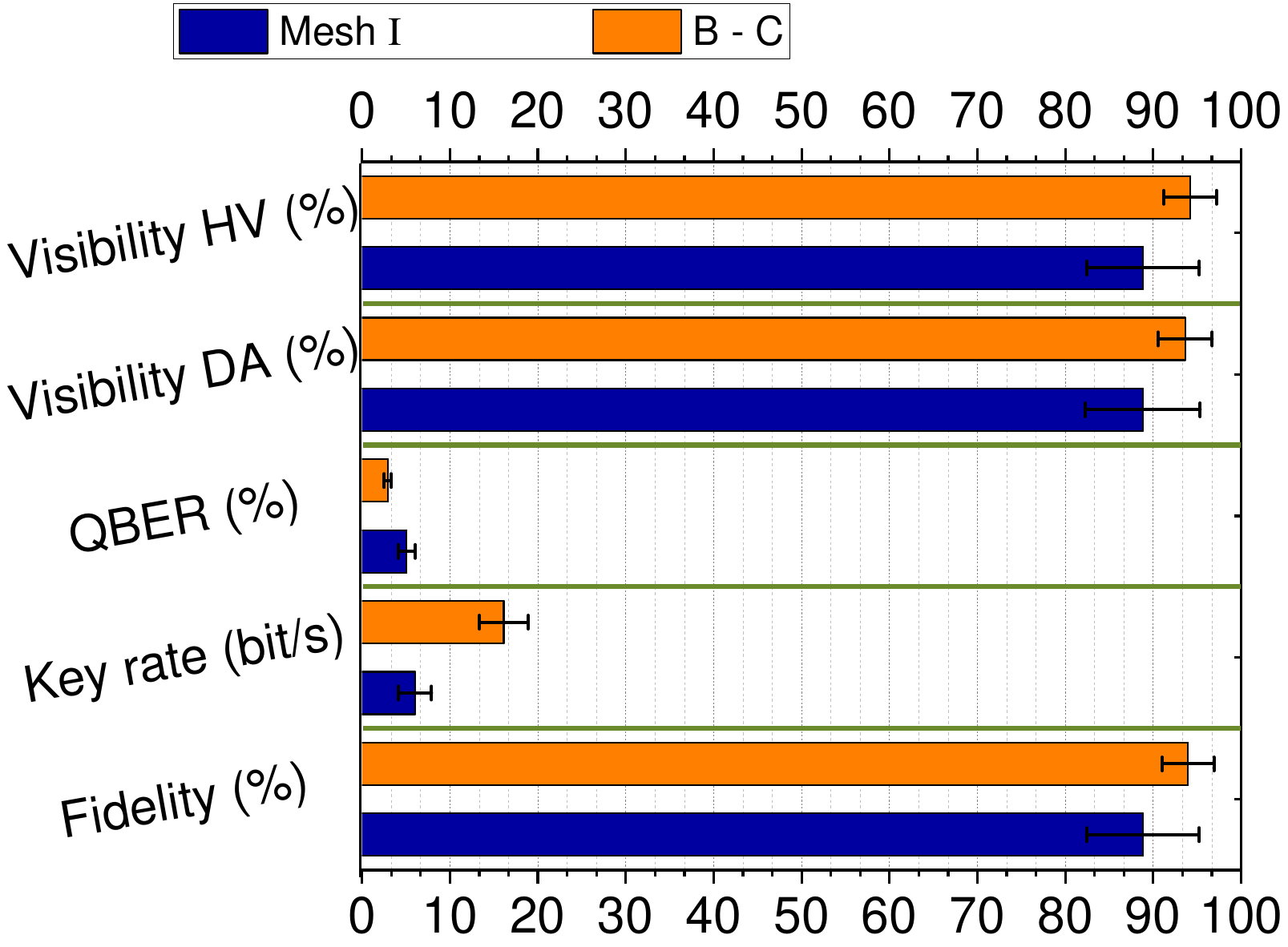}
        \caption{}
        \label{fig:comparsion_new_user}
    \end{subfigure}
    \hspace{0.01\textwidth}
    \begin{subfigure}[t]{0.48\textwidth}
        \centering
        \includegraphics[width=\linewidth]{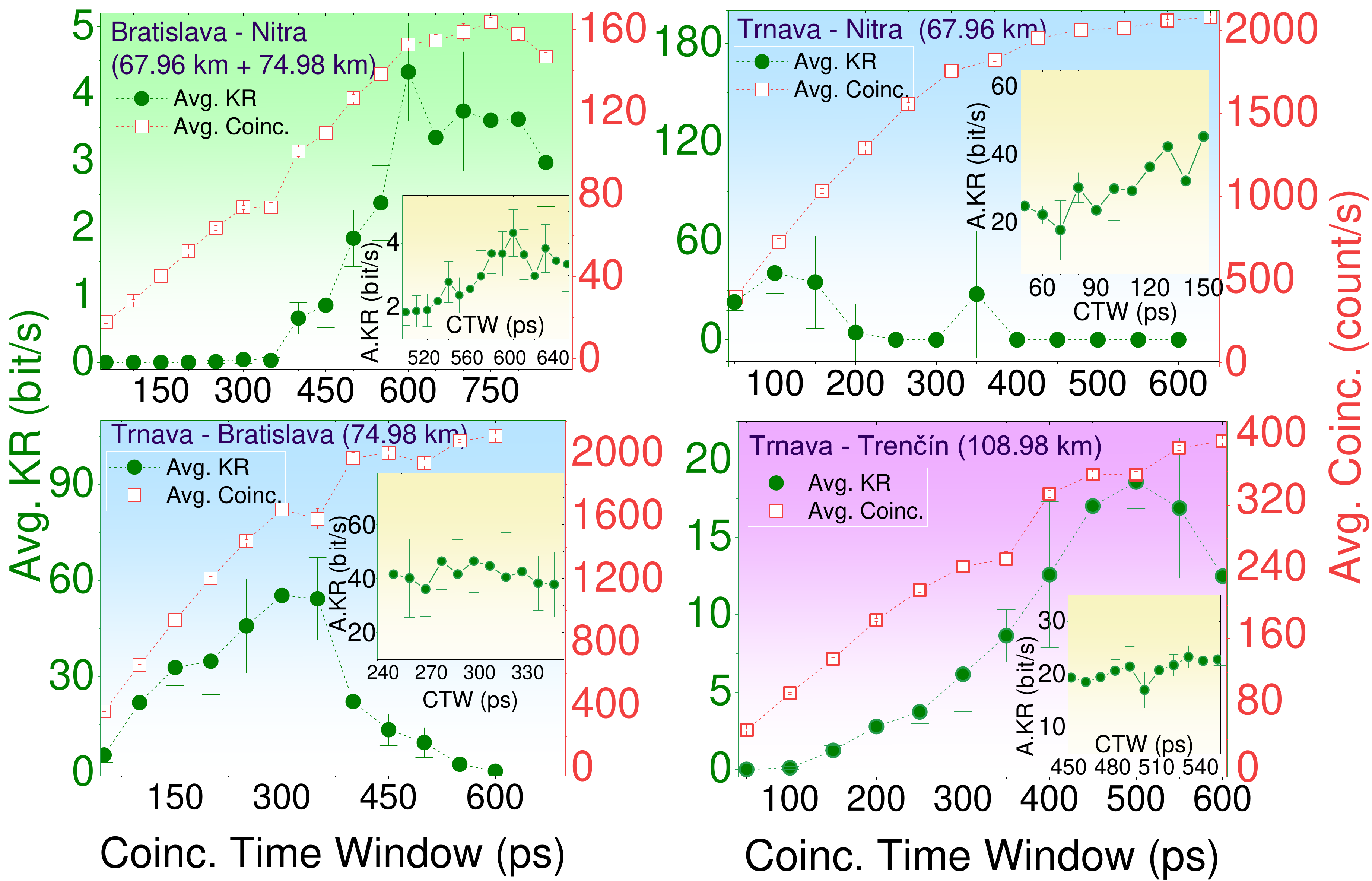}
        \caption{}
        \label{fig:window-analysis}
    \end{subfigure}

    \vspace{0.01cm}

    \begin{subfigure}[t]{0.48\textwidth}
        \centering
        \includegraphics[width=\linewidth]{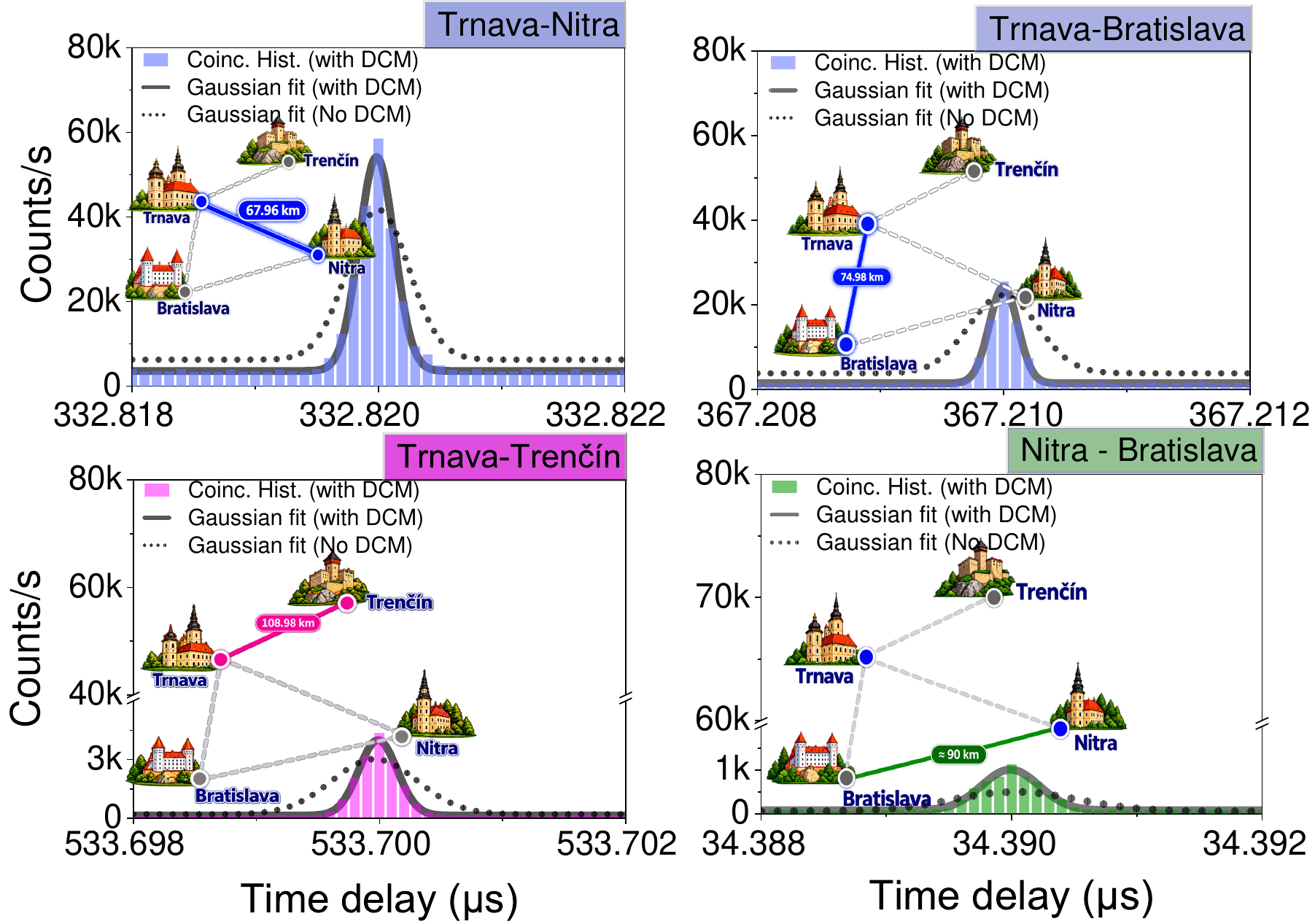}
        \caption{}
        \label{fig:coincidence-histograms}
    \end{subfigure}
    \hspace{0.01\textwidth}
    \begin{subfigure}[t]{0.48\textwidth}
        \centering
        \includegraphics[width=\linewidth]{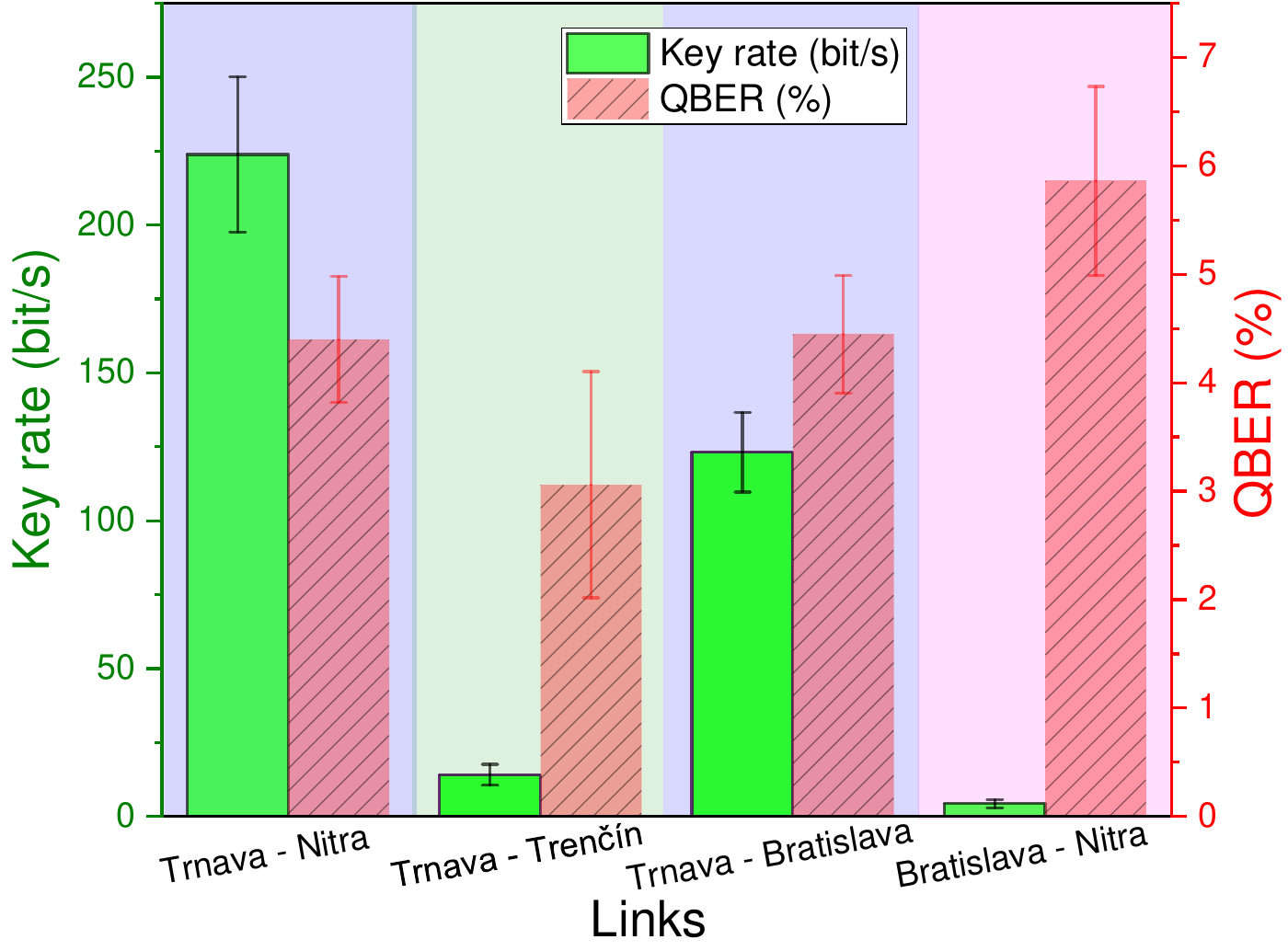}
        \caption{}
        \label{fig:skr-qber}
    \end{subfigure}

    \caption{Experimental results: 
    \subref{fig:comparsion_new_user}) Comparison of the state metrics and QKD performance showing the impact of adding a new user to an established two-user link.
    \subref{fig:window-analysis}) Analysis of the key rate and coincidences as a function of the coincidence-time window
    \subref{fig:coincidence-histograms}) Coincidence histograms with the DCM (bars, 100 ps bins) and their
Gaussian fits (solid). Dotted: Gaussian fits to the histograms recorded without the DCM, whose bars are omitted for readability.
    \subref{fig:skr-qber}) The  key rate (KR) and quantum bit error rate (QBER) obtained in the Mesh II with chromatic dispersion compensation (CDC) .}
    \label{fig:results_deployed_summary}
\end{figure}

For each configuration, we collected data over 24 hours and calculated the average key performance quantities shown in Fig.~\ref{fig:results_deployed_summary}a. We attribute the lower average visibility and fidelity, higher average QBER, and lower average SKR observed in Mesh I to the addition of the new user. These results reinforce the need for an entanglement distribution strategy that minimizes the wavelength load at each node and illustrate the trade-off between key generation rate and the number of nodes, as formalized in Section \ref{subsec:connectivity_matrix} and in ~\cite{panwar2026exact}.

To complete the topology deployed here, we establish the A--D connection using the ITU channel pair 33 (routed to node A) and 35 (routed to node D). Furthermore, we analyzed the influence of the coincidence time window (CTW) on the average key rate over a $25\text{-minute}$ integration period. The optimization results for each link are presented in Fig.~\ref{fig:window-analysis}, which also displays the average net coincidence counts and includes an inset detailing a fine scan of the CTW values around the maximum. In operation, the coincidence time window is not determined solely by the peak width. After the remote time taggers are synchronized through the distributed
clock reference, the window is scanned to maximize the key rate delivered by the live post-processing. The window therefore accounts for accidental coincidences arising from multi-pair emission at the pump power used. On the
lower-rate links the optimum lies close to the HWHM of the peak, while on the
brightest link, Trnava–Nitra, accidentals dominate, and the optimum is narrower
than the peak.

The results obtained in Fig.~\ref{fig:window-analysis} were obtained without compensating for the chromatic dispersion, so it was conducted with a laser pump power of $\approx 14.3~\si{\milli \watt}$ stimulating the source. The obtained optimum values for CTW are shown in Table~\ref{tab:ctw}. With 100 ps bins, binning contributes about $3 \%$ to the fitted widths. The compensated peaks are limited by the synchronization of the remote time
taggers ($\approx 0.3 \si{\nano\second}$ FWHM, against $\approx 60 \si{\pico\second}$ from detector and tagger jitter), so their shape is close to Gaussian. Without compensation, the peaks tend towards the flat-topped filter shape. Since the operational window is chosen by
key-rate optimization, the peak shape does not affect the reported windows or
key rates. Improving the synchronization, rather than further dispersion
compensation, is the main route to narrower windows.
\begin{table}[H]
\centering
\caption{Coincidence time windows without dispersion compensation (laser power 14.3 mW) and with the $-1260~\si{\pico\second \per \nano \meter}$ DCM (laser power 17.3 mW). All values represent the HWHM of the coincidence histograms Gaussian fit shown in Fig.~\ref{fig:coincidence-histograms}
}
\begin{tabular}{lcc}
\toprule
\textbf{Link} & CTW uncompensated (ps) & CTW with DCM (ps) \\
\midrule
A--B & 589 (±0.009) &  336 (± 0.004) \\
A--C & 660 (± 0.011) & 271 (± 0.003) \\
B--C & 1177 (± 0.025) & 541 (± 0.010) \\
A--D & 821 (± 0.012) & 365 (± 0.004) \\
\bottomrule
\end{tabular}
\label{tab:ctw}
\end{table}

\begin{figure}[h]
    \centering
    \includegraphics[width=\linewidth]{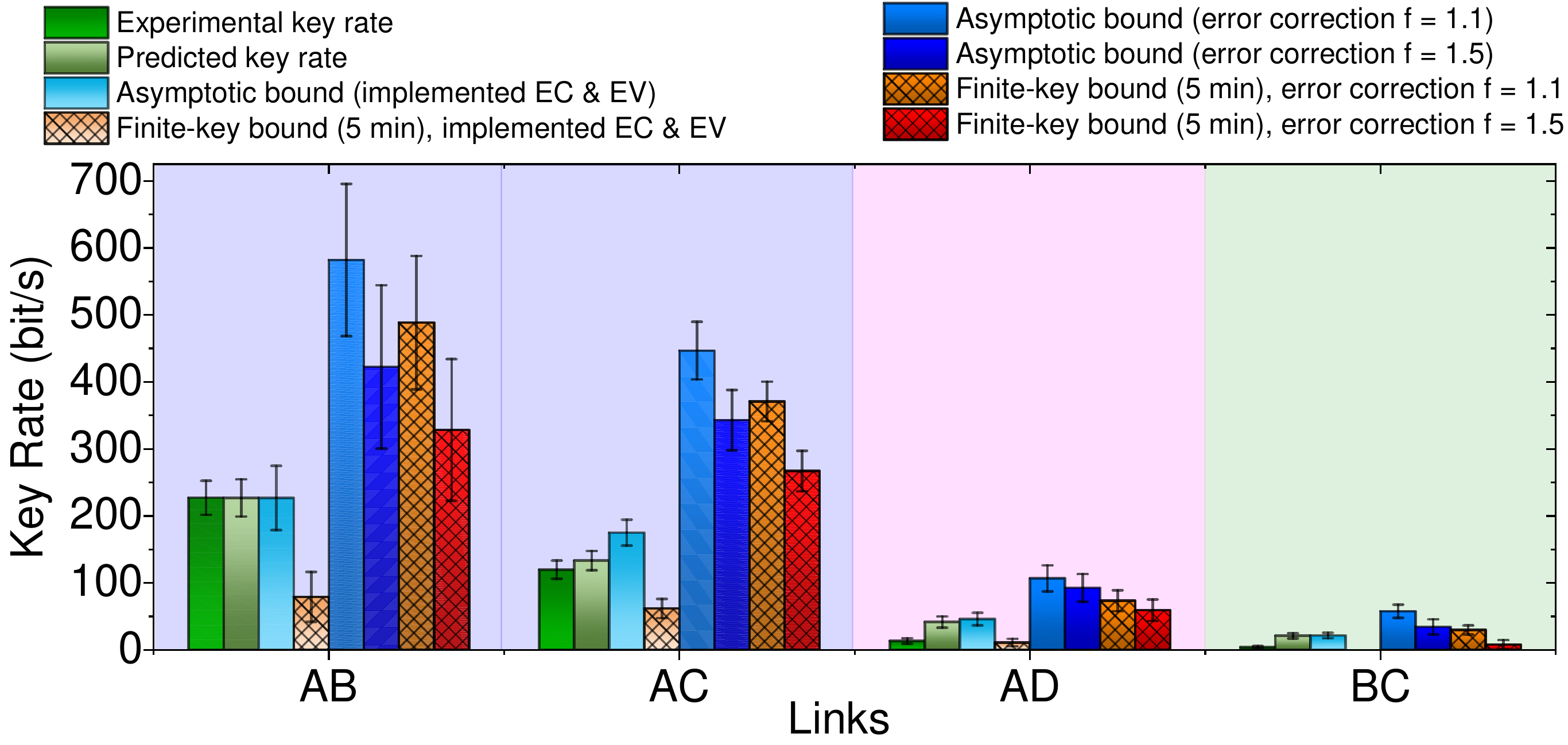}
    \caption{Key rate predictions for the deployed network experimental result. The predicted key rates are determined using Equations~\ref{eq:finite_key_H(e)} and ~\ref{eq:asymptotic_key}. }
    \label{fig:deployedNetwork_comparsion_result}
\end{figure}

We also analyzed the effect of partial chromatic dispersion compensation (CDC), introduced in section~\ref{subsec:Entanglement_generation}. 
Since the broadening depends only on the total dispersion of the pair, the whole compensation can be applied to one photon, here at the central node. A single
$-1260~\si{\pico\second \per \nano \meter}$ module was used for all links. It matches the
Trnava–Bratislava length and leaves a residual dispersion on the longer links, which explains their wider compensated peaks.~\cite{neumann2021nonlocal}. The dispersion compensation module (DCM) utilized in this work (kindly provided by Quantum Optics Jena GmbH) imposes an insertion loss of $\approx 1.5~\si{\decibel}$. To compensate for this minor insertion loss, we increased the laser power to $\approx 17.3~\si{\milli\watt}$, although a higher pump power increases accidental coincidences due to multipair generation. For long links, the narrow CTW helps filter out these events. However, there is an upper limit on the pump power above which the accidental ratio increases the QBER beyond the $11 \%$ threshold. The efficacy of the DCMs is directly observable in the narrowing of the coincidence peak full width at half maximum (FWHM), as compared in Fig.~\ref{fig:results_deployed_summary}c. The SKR and the QBER for the system operating with the optimized condition after CDC are shown in Figure~\ref{fig:results_deployed_summary}d. We can see that for the link A--D, with a total distance of $109~\si{\kilo\meter}$,  the obtained key rates is $\approx 13.5~\mathrm{bit/s}$ while for the link B--C it was $\approx 4.3 ~\mathrm{bit/s}$, with a total propagation distance of $142.94~\si{\kilo\meter}$ and $\approx 32.1 ~\si{\decibel}$ of accumulated loss.

We repeat the benchmark for the deployed network (Fig.~\ref{fig:deployedNetwork_comparsion_result}), using privacy-amplification blocks of $5~\si{\min}$ because of the lower coincidence rates and higher QBER. The distilled key again exceeds the finite-key bound of the implemented EC and EV on all links for which this bound is positive, by a factor of 2.9 on A--B, 1.9 on A--C, and 1.2 on A--D (within one standard deviation). On the longest link, B--C ($142.94~\si{\kilo\meter}$), the finite-key bound of the implemented scheme vanishes, so none of the distilled key can be certified, whereas error correction with $f = 1.1$ would still yield $29.8 \pm 6.8$\,bit/s. On the remaining links, error correction would increase
the finite-key bound by a factor of 4.1--6.6.

\section{Conclusions}\label{sec:Conclusion}

We have demonstrated an inter-city entanglement-based QKD network connecting four cities in western Slovakia, in which a trusted-node-free, fully meshed sub-network of three cities and a fourth node connected through the central node are operated simultaneously, with the complete key distillation and delivery chain running in real time. The deployed system is a commercial evaluation solution developed within the EuroQCI ecosystem as part of SkQCI. With partial non-local chromatic dispersion compensation, it delivers $13.5~\mathrm{bit/s}$ of final key over $109~\si{\kilo\meter}$ and $4.3~\mathrm{bit/s}$ over $142.94~\si{\kilo\meter}$, with $32.1~\si{\decibel}$ of accumulated loss between two cities without a direct fiber link. We also measured the effect of adding a user to an existing mesh.

Benchmarking the implemented post-processing against asymptotic and finite-key security bounds shows that the key length set by the heuristic privacy amplification can exceed the finite-key bound for short blocks, so finite-size corrections must be included before the delivered key can be considered secure. Additionally, we identify that replacing the detect-and-discard key reconciliation with error correction would increase the finite-key rate by a factor of 2.1--6.8 on the deployed links. 

Finally, we introduced a matrix formalism for passive wavelength routing that verifies the connectivity of a given QNP, providing the basis for the resource laws derived in our companion work~\cite{panwar2026exact}, which extends this approach to a graph-theoretic treatment of arbitrary networks.

\section*{Acknowledgment}

This work was carried out within the project skQCI -- Slovak Quantum Communication Infrastructure, funded by the European Union under the Digital Europe Program, call
\newline
\mbox{DIGITAL-2021-QCI-01-DEPLOY-NATIONAL}, with 50\,\% national co-financing from the Recovery and Resilience Plan of Slovakia, as part of the European Quantum Communication Infrastructure (EuroQCI) initiative. We thank Du\v{s}an Lacko and Peter Lacko for their work on early post-processing software tests and for discussions on the network back-end. We also thank David Polzoni for discussions on QKD post-processing and Younes Benamara for technical assistance during deployment.

We thank the institutions hosting our nodes: the Faculty of Materials Science and Technology in Trnava of the Slovak University of Technology in Bratislava, the Slovak University of Agriculture in Nitra, and Alexander Dub\v{c}ek University of Tren\v{c}\'{\i}n; the Slovak Academic Network (SANET) for dedicated intra-city fibers and IP connectivity, and the Computing Center of the Slovak Academy of Sciences for optical connectivity on the Bratislava campus. 
The authors would like to thank the commercial partners, Quantum Optics Jena GmbH and Zero-Third GmbH, for their assistance and discussions. 
We acknowledge the use of Claude for grammar checking and assitance in illustration in Figures ~\ref{fig:network_slovak_map} and \ref{fig:topo_ABC}. The authors reviwed, edited, and maintian full responsability for the final content. 
Views and opinions expressed are those of the authors only and do not necessarily reflect those of the European Union or the European Commission. Neither the European Union nor the granting authority can be held responsible for them.

\section*{Funding}
Digital Europe Programme (101091548, skQCI); 
Recovery and Resilience Plan of Slovakia (skQCI); 
Horizon 2020 Marie Sk\l{}odowska-Curie COFUND Action (945478, SASPRO2, fellowship 1156/01/01, BIQuaN); 
Fellowships for excellent researchers R2-R4 (09I03-03-V04-00777, QENTAPP). 

\section*{Disclosures} 
The authors declare no conflicts of interest. 

\section*{Data availability} Data underlying the results presented in this paper are not publicly available at this time but may be obtained from the authors upon reasonable request. 

\bibliography{references,supplementary_refs}

\clearpage
\appendix
\setcounter{section}{0}
\setcounter{figure}{0}
\setcounter{table}{0}
\setcounter{equation}{0}
\renewcommand{\thesection}{S\arabic{section}}
\renewcommand{\thefigure}{S\arabic{figure}}
\renewcommand{\thetable}{S\arabic{table}}
\renewcommand{\theequation}{S\arabic{equation}}
\renewcommand{\theHsection}{S\arabic{section}}
\renewcommand{\theHfigure}{S\arabic{figure}}
\renewcommand{\theHtable}{S\arabic{table}}
\renewcommand{\theHequation}{S\arabic{equation}}
\lstset{style=optica}

\begin{center}
  {\Large\bfseries Supplement 1}\\[4pt]
  {\large A deployed four-city entanglement-based QKD network with real-time key delivery, benchmarked against finite-key security bounds: supplemental document}
\end{center}
\vspace{1em}

\section{Matrix formalism for passive wavelength routing: examples}
Here we provide two examples of network topologies and apply the matrix formalism introduced in Sec. 2.5 to describe them.
Consider a four-user network with six wavelength channels ($N = 4$, $M = 6$), in which each channel can be split at most once by a standard $1 \times 2$ BS. Tables~\ref{tab:no_UC} and~\ref{tab:UC_users} show two fully meshed configurations generated by the algorithm. In Table~\ref{tab:no_UC}, all channels are split, and every pair of users shares two entangled wavelength pairs, so applying the connectivity rule of main-text Eq.~(7) gives  $\mathbf{C}_{\text{non-UC}} = 2(\mathbf{J}_4 - \mathbf{I}_4)$, and every link is a premium link. In Table~\ref{tab:UC_users}, only the channels $\lambda_4$ to $\lambda_6$ are split, and every pair of users shares exactly one entangled pair, so $\mathbf{C}_{\text{UC}} = \mathbf{J}_4 - \mathbf{I}_4$ as required by main-text Eq.~(9). The UC configuration thus achieves full connectivity with half of the beam splitters, and the unsplit channels reach the users without splitting loss. These four-user examples describe the complete laboratory topology, not the four-city deployed Mesh~II.

\begin{subequations}\label{eq:connect_4user_matrix}
\begin{align}
    \mathbf{C}_{\text{non-UC}} & = \begin{pmatrix}
        0 & 2 & 2 & 2 \\
        2 & 0 & 2 & 2 \\
        2 & 2 & 0 & 2 \\
        2 & 2 & 2 & 0
    \end{pmatrix} \: \label{eq:non-UC_C_matrix} \\[1.em]
    \mathbf{C}_{\text{UC}}& = \begin{pmatrix}
        0 & 1 & 1 & 1 \\
        1 & 0 & 1 & 1 \\
        1 & 1 & 0 & 1 \\
        1 & 1 & 1 & 0
    \end{pmatrix} \label{eq:UC_C_matrix}
\end{align}
\end{subequations}

\begin{table}[htpb]
\centering
\begin{minipage}{0.45\textwidth}
\centering
\caption{QNP table with not UC users.}
\begin{tabular}{c|cccc}
 & $U_1$ & $U_2$ & $U_3$ & $U_4$  \\
\hline
$\lambda_1$ & {\color{blue}1} & 0 & 0 & {\color{blue}1} \\
$\lambda_2$ & {\color{orange}1} & 0 & {\color{orange}1} & 0 \\
$\lambda_3$ & {\color{red}1} & {\color{red}1} & 0 & 0 \\
\hdashline
$\lambda_4$ & 0 & 0 & {\color{red}1} & {\color{red}1} \\
$\lambda_5$ & 0 & {\color{orange}1} & 0 & {\color{orange}1} \\
$\lambda_6$ & 0 & {\color{blue}1} & {\color{blue}1} & 0 \\
\end{tabular}
\label{tab:no_UC}
\end{minipage}\hfill
\begin{minipage}{0.45\textwidth}
\centering
\caption{QNP table with UC users.}
\begin{tabular}{c|cccc}
 & $U_1$ & $U_2$ & $U_3$ & $U_4$  \\
\hline
$\lambda_1$ & 0 & {\color{blue}1} & 0 & 0 \\
$\lambda_2$ & 0 & 0 & {\color{orange}1} & 0 \\
$\lambda_3$ & 0 & 0 & 0 & {\color{red}1} \\
\hdashline
$\lambda_4$ & {\color{red}1}  & {\color{red}1} & 0 & 0 \\
$\lambda_5$ & {\color{orange}1} & 0 & 0 & {\color{orange}1} \\
$\lambda_6$ & {\color{blue}1} & 0 & {\color{blue}1} & 0 \\
\end{tabular}
\label{tab:UC_users}
\end{minipage}
\end{table}
\section{Connectivity test and graph representation}

The QNP matrix Q is the binary M × N wavelength–user assignment of main-text Sec. 2.5. Its rows correspond
to wavelength channels in spectral order, and conjugate channels have reverse row indices (main-text Eq. (2)).
The connectivity matrix and the UC full-mesh criterion are given in the main text. (7) and (9);  The direct NumPy implementation is: 
\begin{lstlisting}
import numpy as np

def connectivity_matrix(network):
    nw_90c = np.rot90(network, k=-1)
    C = nw_90c @ network
    return C
\end{lstlisting}

Here, network is the binary array Q, and clockwise rotation aligns each row with its conjugate channel. The first test returns the connection multiplicity for every ordered pair of users. The second accepts only if every distinct pair is connected once and there is no repeated links. In particular, Notebook 1 verifies the UC routing table of supplementary Table~S2 and recovers main text Eq. (S1b). The earlier complete-graph QNP design was presented in the Slovak QCI proceedings work of Aktas \emph{et al.}~\cite{AktasCAMA2025}.

For the counting procedure, it is convenient to treat the $N$ users as the vertices of a complete graph $K_N$~\cite{Diestel2017}. Its required unordered edge set is
\begin{equation}
E(K_N)=\bigl\{\{u_i,u_j\}:1\le i<j\le N\bigr\},
\qquad |E(K_N)|=\binom{N}{2}.
\label{eq:edges}
\end{equation}
An assignment is accepted precisely when its conjugate-pair routes cover every edge in this set once, with no self-links. This is the graph interpretation of the test above, rather than an additional physical assumption. The notebooks enumerate the restricted one-sided assignment model below; they do not implement every user-selected splitting subset of the more general QNP design procedure.
\subsection{One-sided raw assignment ensemble}

The probabilities below apply to a specified proposal ensemble, not to a uniform distribution over distinct physical QNP matrices. Let $L=M/2$ denote the number of labeled conjugate-wavelength pairs. In each layer, one unsplit idler channel is assigned to a single user $s$, and the conjugate signal channel is proposed to an unordered multiset $I$ of $r$ user labels, where
\begin{equation}
 r=\left\lceil\frac N2\right\rceil.
\label{eq:r}
\end{equation}
This choice of $r$ defines the enumeration model and is not a universal fan-out limit. In the $N=3,4$ cases in Table~S3, $r=2$, so a signal wavelength reaches at most two distinct users, consistent with one $1\times2$ splitter. The physical orientation agrees with the main text, Sec.~2.2: the idler side is unsplit, and the signal side may be split.

Multiset labels may repeat. Before constructing the binary QNP matrix, the labels in $I$ are reduced to their distinct support $T=\operatorname{supp}(I)$. A proposed layer is rejected if $s\in T$ because it would route both conjugate photons to the same user. Otherwise, the layer contributes a star with edges
\begin{equation}
\operatorname{star}(s,T)=\bigl\{\{s,t\}:t\in T\bigr\},
\qquad \varnothing\ne T\subseteq\{1,\dots,N\}\setminus\{s\}.
\label{eq:star}
\end{equation}
Repeated edges across different layers are also rejected. Every raw multiset is counted as a distinct equiprobable proposal even if several proposals produce the same binary matrix; this distinction determines the probabilities.

There are $N$ anchor choices and $\binom{N+r-1}{r}$ unordered size-$r$ multisets in each of the $L$ labeled layers. Thus, writing $A(N,M)$ for the number of accepted raw proposals, the total and success fraction are
\begin{equation}
D(N,M)=\left[N\binom{N+r-1}{r}\right]^L,
\qquad P(N,M)=\frac{A(N,M)}{D(N,M)}.
\label{eq:denom}
\end{equation}
Odd $M$ is excluded because every layer contains exactly two conjugate channels.

\section{Exact integer enumeration}

It is unnecessary to loop over all $D(N,M)$ raw proposals to obtain the numerator. For a nonempty recipient set $T$ with $k=|T|\le r$, the number of size-$r$ multisets having support \emph{exactly} $T$ equals the number of ways to partition $r$ into $k$ positive multiplicities:
\begin{equation}
w_r(T)=\binom{r-1}{k-1},\qquad 1\le k\le\min(r,N-1).
\label{eq:weight}
\end{equation}
Let $F_\ell(H)$ count $\ell$-layer raw proposals with no duplicated or self-edges whose union is exactly the edge subset $H\subseteq E(K_N)$. Start with $F_0(\varnothing)=1$ and $F_0(H)=0$ for nonempty $H$. For every center $s$, allowed $T$, and previously attained $H$ disjoint from the new star, apply
\begin{equation}
\begin{split}
F_{\ell+1}\bigl(H\cup\operatorname{star}(s,T)\bigr)
&\mathrel{+}=F_\ell(H)\binom{r-1}{|T|-1},\\[-2pt]
&\hspace{-1pt}H\cap\operatorname{star}(s,T)=\varnothing.
\end{split}
\label{eq:recurrence}
\end{equation}
The exact number of accepted raw proposals is
\begin{equation}
A(N,M)=F_L\bigl(E(K_N)\bigr).
\label{eq:numerator}
\end{equation}
The disjointness test excludes repeated links and the restriction $s\notin T$ excludes self-links. Hence, the recurrence counts exactly the configurations accepted by the QNP test in the main text, Eq.~(9).

\noindent\textbf{Algorithm S1 (exact counting).} The following standalone Python implementation uses integer bit masks for the edge sets. Notebook~1 checks small cases by direct enumeration; Notebook~2 uses the same recurrence to reproduce the full table and figure.
\begin{lstlisting}
from collections import defaultdict
from itertools import combinations
from math import comb

def exact_probability(N, M):
    if M % 2:
        raise ValueError("M must be even")
    L, r = M // 2, (N + 1) // 2
    edges = list(combinations(range(N), 2))
    index = {e: j for j, e in enumerate(edges)}
    stars = []
    for s in range(N):
        others = [t for t in range(N) if t != s]
        for k in range(1, min(r, N-1) + 1):
            for T in combinations(others, k):
                mask = sum(1 << index[tuple(sorted((s,t)))]
                           for t in T)
                stars.append((mask, comb(r-1, k-1)))
    counts = {0: 1}
    for _ in range(L):
        following = defaultdict(int)
        for used, value in counts.items():
            for star, multiplicity in stars:
                if not (used & star):
                    following[used | star] += value * multiplicity
        counts = following
    accepted = counts.get((1 << len(edges)) - 1, 0)
    total = (N * comb(N+r-1, r)) ** L
    return accepted, total  # probability = accepted / total
\end{lstlisting}

\begin{table}[H]
\centering
\caption{Exact acceptance probabilities for a uniquely connected complete QNP assignment in the one-sided raw multiset model. Accepted counts $A$ and total counts $D$ are integers; $P=A/D$ is rounded for display.}
\label{tab:S1}
\small
\begin{tabular}{@{}ccrrr@{}}
\toprule
Users $N$ & Channels $M$ & Accepted $A(N,M)$ & Total $D(N,M)$ & $P(N,M)$\\
\midrule
3 & 4 & 12 & 324 & 0.037037\\
3 & 6 & 48 & 5,832 & 0.008230\\
4 & 6 & 48 & 64,000 & 0.000750\\
4 & 8 & 2,880 & 2,560,000 & 0.001125\\
4 & 10 & 23,040 & 102,400,000 & 0.000225\\
4 & 12 & 46,080 & 4,096,000,000 & 0.00001125\\
\bottomrule
\end{tabular}
\end{table}

\begin{figure}[H]
\centering
\includegraphics[width=0.61\linewidth]{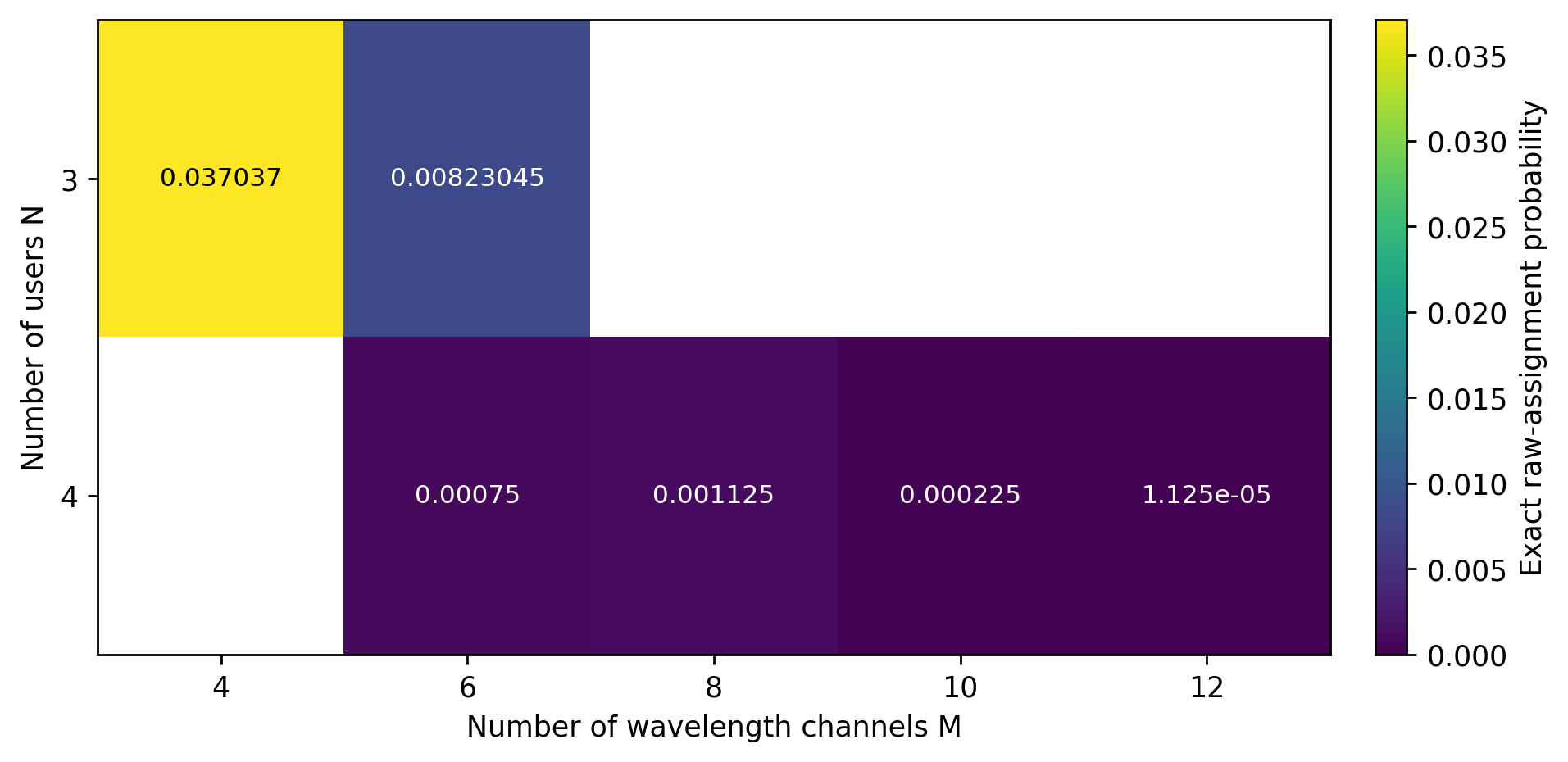}
\caption{Probability $P(N,M)$ for the pairs tabulated in Table~S3, calculated from exact integer counts. Blank positions correspond to omitted parameter pairs; all blanks in the displayed grid have zero acceptance probability in this model since a valid assignment must satisfy $L\le\binom N2\le rL$.}
\label{fig:S1}
\end{figure}

\end{document}